\documentstyle[proceedings,psfig]{crckapb}

\newcommand{\x}{\mbox{$\times$}}
\newcommand{\Mo}{\mbox{M$_{\odot}$}}
\newcommand{\Zo}{\mbox{Z$_{\odot}$}}

\begin{opening}

\title{TIDAL DWARF GALAXIES}

\author{P.-A. DUC }
\institute{ESO, 
         Karl-Schwarzschild Strasse 2\\
         D-85748 Garching bei M\"unchen, Germany\\ 
$<$pduc@eso.org $>$}
\author{I.F. MIRABEL }
\institute{CEA, SAp, C.E. Saclay\\
           91191 Gif/Yvette Cedex, France, \&
Instituto de Astronom\'\i a y F\'\i sica del Espacio, Argentina\\ 
$<$mirabel@discovery.saclay.cea.fr $>$}

\end{opening}

\runningtitle{TIDAL DWARF GALAXIES}

\begin{document}

\begin{abstract}
We review the observational evidences for tidal dwarf galaxies, a class of 
small galaxies formed out of the tidal debris of collisions between 
massive galaxies. Tidal dwarfs are found far from the interacting parent 
galaxies, associated to massive clouds of atomic hydrogen located at the tip of 
long tidal tails. These newly formed galaxies are 
among the best cases for the study of galaxy formation in the nearby Universe. 
\end{abstract}

\section{Introduction}
Most studies of interacting/merging systems focus on their central
regions, which are dramatically affected by the collision.
In particular the gas often looses angular momentum and sinks into the
 galactic cores,
triggering  nuclear starbursts and the formation of young star
clusters.
On the other hand tidal forces may pull out from the outer regions into
 intergalactic space
stars and interstellar gas shaping  rings, bridges, plumes and tails.
The amount of matter lost during that outflow can be
as large as one third of the mass in the pre-encounter disks.
In interacting systems the bulk of the atomic hydrogen is in fact
located outside the galaxy bodies 
 (see review  by F. Combes in this volume and
examples below). \\

Nevertheless until recently  only  few studies have been devoted to the
 properties of tidal features  (e.g. Wallin, 1990; Schombert et al., 1990).
 Tails are basically used as tracers of past 
 minor/major mergers. Their
shapes are useful to constrain the parameters in numerical models of the
 collision, including the mass of dark
matter in the parent galaxies (Dubinski et al., 1996).
The interest in these collisional debris was revived when it
was discovered that they host active star forming regions
(Schweizer, 1978; Mirabel et al., 1992) and are actually a
nursery of small galaxies, the so-called ``tidal dwarf galaxies''
 (hereafter TDGs). 
New  optical and HI observations have  shown that
TDGs  actually form a class of ``recycled'' objects with some characteristics
similar to the more classical dwarf irregulars (dIrrs) and blue
compact dwarf galaxies (BCDGs). Here we review prototype
interacting systems that exhibit the formation of tidal dwarfs,
detail the general properties of TDGs and give some hints for
their origin and future evolution.

\section{Case studies}

\begin{figure}
%\centerline{\psfig{file=fig1.ps,width=\textwidth}}
\caption{Tidal dwarf galaxies in four interacting systems.
They are seen at the end of optical tidal tails or HI plumes (black
contours) and indicated by arrows. The optical images are clock-wise from
 the Digital Sky Survey, Duc (1995),
Duc \& Mirabel (1997) and Duc \& Mirabel (1994). The HI maps  are from
Hibbard et al. (1997), Hibbard et al. (1994), Malphrus et al. (1997) and
 Duc et al. (1997)}
\label{fig:compose}
\end{figure}

\begin{figure}
%\centerline{\psfig{file=fig2.ps,width=6cm}}
\caption{Optical image of the disk-disk system NGC 2992/93 with the 
H$\alpha$ emission superimposed in white. Note how star formation as traced
by the ionized gas is enhanced simultaneously in the galaxy body and
at the tip of the tidal tail.}
\label{fig:N2992}
\end{figure}

We have carried out multi-wavelength observations of several interacting
systems in the nearby Universe.
Optical imaging and spectroscopy 
were obtained with the CFHT at Mauna Kea and the ESO/NTT at la Silla
Observatory;
near-infrared imaging at the ESO/MPI 2.2m, and HI at the VLA.
HI data come either from our own observations, or were kindly provided
to us by other groups.

Fig.~\ref{fig:compose} and Fig.~\ref{fig:N2992} present different examples
 of interactions: spiral-spiral
collisions (NGC 4038/39, ``The Antennae''; NGC 2992/93), complete merger
 between spirals (NGC 7252) and
  encounters involving early-type galaxies (Arp 105, ``The Guitar galaxy'';
NGC 5291). Long tidal tails are clearly seen
 emanating from the parent galaxies. At their tip, at  distances up to
 100~kpc from the nuclei,  small irregular objects are found with absolute
magnitudes typical of dwarf galaxies. They host blue compact
clumps that also show up in maps of
 the ionized gas (see Fig.~\ref{fig:N2992}).

\begin{figure}
\centerline{\psfig{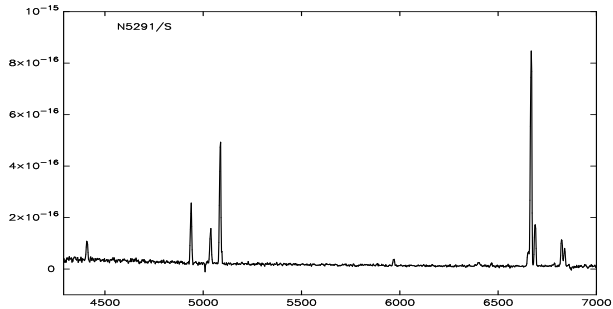}}
\caption{Optical spectrum of one of the tidal dwarfs near NGC 5291
showing emission lines typical of HII regions. From
 Duc \& Mirabel (1997)}
\label{fig:spectra}
\end{figure}

 The spectra of the optical condensations exhibit emission  lines, typical
of HII regions ionized by massive OB stars younger than 10~Myrs
(e.g. Fig.~\ref{fig:spectra}).
Given the time scale for the formation of clumps in  tails --
 typically 1~Gyr  --,  the stars at the tip of the antennae must have been 
 born in situ. Therefore interactions not only trigger
star formation  in the main body of the parent galaxies but also
 far from the nuclei in the remote tidal features.

 The contours of the HI column density
are superimposed  on the optical images in Fig.~\ref{fig:compose}.
 It is clear that the central regions of the parent galaxies
contain little atomic gas, whereas the optical  tails,
and especially the tidal dwarfs at their tip, are associated
with  HI clouds as massive as  $6 \x 10^{9}$~\Mo. Similar HI
distributions were observed by Hibbard \& van Gorkom (1996) in
several other interacting systems.
 On the other hand the molecular gas tends to concentrate in the
central regions fueling the nuclear starburst of interacting galaxies
 (e.g. Young \& Scoville, 1991).
 Such a spatial segregation of the different gas components is clearly 
seen in Arp 105 (Duc et al., 1997;  Fig.~\ref{fig:compose}).

\section{Properties of tidal dwarf galaxies}

Table~1 summarizes the statistics for the main properties of the 20 TDGs
 sofar studied.

\begin{table}
\caption{Integrated properties of tidal dwarf galaxies}
\begin{tabular}{llcccc}
Properties & Units &  Mean & Min & Max \\
\hline
Absolute Nlue Magnitude & mag  & -14.8 & -12.1 & -18.8 \\
B-V color index & mag  & 0.3 & 0.0 & 0.7 \\
%    &     & 13 & 0.3 & 0.2 & 0.5 \\
%    &     &  4 & 0.4 & 0.2 & 0.7 \\
Star Formation Rate & log(\Mo/year)  & -1.1 & -3.6 & 0.3 \\ 
HI Mass & $10^9$~\Mo  & 1.6 & 0.2 & 6.0 \\
O Abundance & 12+log(O/H)  & 8.5 & 8.3 & 8.6 \\  
\hline
\end{tabular}
\end{table}

\subsection{Stellar populations and gas content}
\begin{figure}
\centerline{\psfig{file=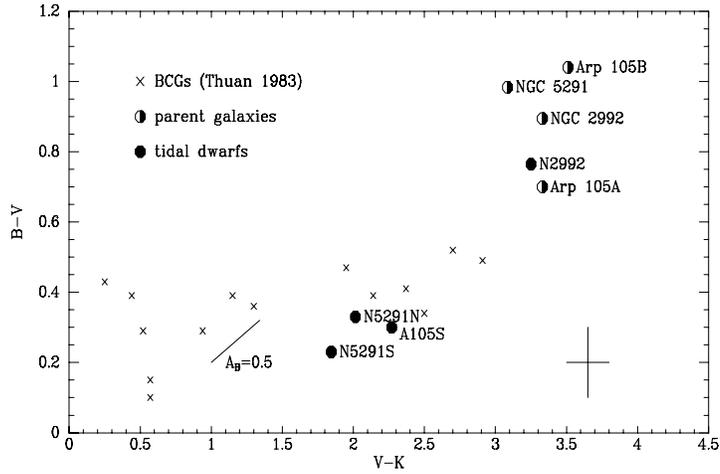,width=10cm}}
\caption{Color-color diagram of tidal dwarfs. For reference, the colors
of the outer regions of the parent galaxies are also indicated and a
sample of blue compact dwarfs  has been added. The cross at the lower
right corner indicates typical error bars.}
\label{fig:col}
\end{figure}

Tidal dwarf galaxies may be made of two stellar components:
young stars,  formed from the recent collapse of expelled
HI clouds, and possibly an older star population coming from the
 disk of their  parent galaxies.
Using our aperture photometry and spectroscopic data,
 we could estimate
the relative proportion of both populations and conclude that TDGs
actually split into two categories 1) extremely young objects, most 
probably forming
their first generation of stars (e.g.: the dwarfs around NGC~5291,
Duc \& Mirabel, 1997), with high star formation rates equivalent to those
observed in blue compact dwarf galaxies, and 2) galaxies dominated by an old
stellar population originally from the disk of their progenitors, and
 that look like dwarf irregulars (e.g. the galaxy North of NGC 2992,
 Fig.~\ref{fig:N2992}). Both type of galaxies
  separate on an optical/near infrared color-color diagram, as
 shown in Fig.~\ref{fig:col}. The optical B-V colors of the ``young''
TDGs  and BCDGs are similar. The near-infrared V-K color index of TDGs
appears though to be redder on average, which could be due to a difference
in metallicity between both classes of objects.
The colors of the ``old'' TDGs are similar to these of 
the outer parts of their parent galaxies. 

In all these objects, the equivalent width of the optical Balmer
lines indicates that the current star--forming episode is 
younger than 10 Myrs. Important HI reservoirs -- between 
$5 \x 10^{8}$~\Mo~ and $5 \x 10^{9}$~\Mo -- can sustain the
star formation for  several Gyrs.

The molecular gas content of tidal dwarfs is still largely unknown.
Smith \& Higdon (1994)  failed to detect any CO emission in the tails
of a few interacting systems. CO was also reported to be very weak in 
 blue compact dwarf galaxies despite their high star formation
rate. This was interpreted as a metallicity effect.
However TDGs seem to be quite metal rich systems in which CO
should be easier to detect. 

\begin{figure}
\centerline{\psfig{file=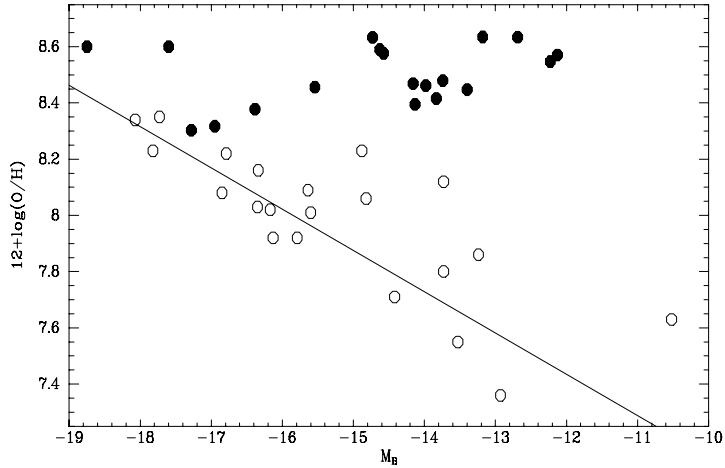,width=10cm}}
\caption{Oxygen abundance vs absolute blue magnitude for our sample of tidal 
dwarfs (black points) and a sample of isolated dwarf 
 galaxies (open points; from Richer \& McCall, 1995).}
\label{fig:abun}
\end{figure}

\subsection{Metallicity}
Fig.~\ref{fig:abun} shows the oxygen abundance vs absolute magnitude of a
 sample of TDGs and nearby dIrrs. The abundances have been estimated in
the ionized gas from the  [OIII]/H$\beta$ line ratio. Their uncertainties
 are discussed in Duc \& Mirabel (1997). 
 Clearly TDGs are more metal rich than classical
dwarfs of the same luminosity. They have metallicities $\Zo/3$ on average,
 a value  that is typical
 of the outer regions of spirals. They do not follow the  correlation
found for field dwarf and giant galaxies between luminosity
(hence mass) and metallicity. Being ``recycled'' objects, formed from
pre-enriched material, tidal dwarfs got as an heritage from their parents 
this relative high metal content.

\subsection{Dynamics}
\begin{figure}[h]
\centerline{\psfig{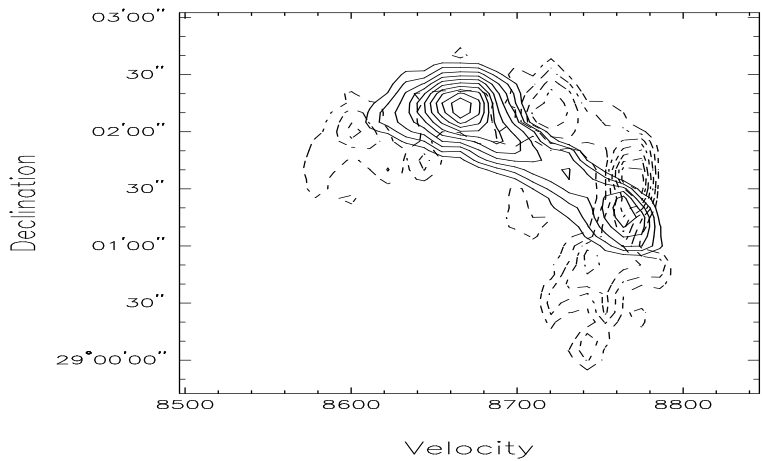}}
\caption{Position-Velocity diagram of the HI
northern tidal tail in Arp~105. Two components were identified:  the
 expanding HI 
tidal tail (dashed contours) and a kinematically decoupled,
 possibly rotating, component (continuous contours) associated with the
 formation of a
 tidal dwarf at the tip of the tail; adapted from
 Duc et al. (1997) }
\label{fig:vel_A105}
\end{figure}

\begin{figure}[h]
\centerline{\psfig{file=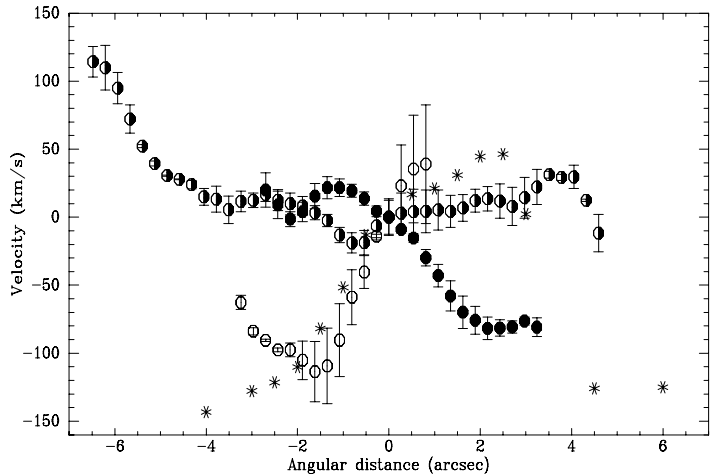,width=10cm}}
\caption{Velocity profiles of the ionized gas for several tidal dwarfs;
adapted from Duc \& Mirabel (1997)}
\label{fig:vel}
\end{figure}

Little is still known about the internal dynamics of tidal dwarfs. First
indications are that the most massive TDGs may be gravitationally
bound (Hibbard et al., 1997; Malphrus et al., 1997). Hints for rotation
  were found in the HI cloud associated with a TDG in Arp~105
 (see Fig.~\ref{fig:vel_A105}).
Furthermore, strong velocity gradients of the ionized gas have been found
 in several objects, which suggests rotation too (Fig.~\ref{fig:vel}).
Therefore,  some objects  in tidal tails may  already be dynamically
independent. Further 3D kinematical studies would be necessary to verify this
 assertion and estimate their dark matter content, predicted to be low
in numerical simulations (Barnes \& Hernquist, 1992).\\

\section{Origin and evolution of tidal dwarf galaxies}

\subsection{Ingredients for the formation of TDGs}
Obviously star-forming tidal dwarf galaxies only form after an encounter
between gas-rich galaxies. The nature of the collision is  
a priori also important.
One should  differentiate disk-disk collisions from collisions
involving  early type galaxies. In the first case, tidal
 features are made out of expelled gas and stars. In the second case,
 stellar tails are 
difficult to form because of the  pressure-supported nature of
the stellar dynamic of ellipticals, whereas the HI component, mostly of
 external origin and often supported
by rotation, is easier to disrupt. Pure HI tails will then be formed.
Finally there are instances where  HI and stellar tails
have different positions and extensions. The initial properties
of the resulting TDGs will depend on their location. This explains
why two categories of TDGs have been identified.

\subsection{Models for the formation of TDGs}
Models for the formation of TDGs put forward
two mechanisms: a local dynamical instability in the old stellar
populations of tidal tails, followed by accretion of gas (Barnes \& Hernquist,
 1992) or the collapse of a supermassive cloud triggering precipitous star
formation activity (Elmegreen et al., 1993). Because of the
variety of the stellar and gas content of TDGs one could argue that
both mechanisms play a role: the stellar scenario would apply for 
TDGs born in HI+stellar tails; the gas scenario for TDGs born in pure HI
 tails. However it seems that even for TDGs belonging to the first
category,  the HI masses and resulting potential well are large enough
 to  trap the old tidal star
 population. As a consequence, despite the diversity of
initial environments, all TDGs may have been formed by the same mechanism.

What causes the collapse of tidally expelled gas clouds is still unclear.
Other environmental effects such as 
ram pressure by the intergalactic medium may play a role. 
Strong asymmetries of the HI distribution in some interacting systems
suggest some kind of compression at the interface with the IGM.
Hawarden \& Chaytor (1996) find in NGC~5291 a possible expansion of the x-ray 
emission along the HI tail and its associated HII regions.
This hypothesis should be further investigated and modelized.
In x-ray clusters, besides tidal forces, ram pressure could even have
 taken part in
the pulling out of the HI clouds from spirals known to be HI deficient.
Instances of enhanced star formation in stripped clouds  have been
presented by J. Kenney (this volume).

\subsection{Survival of TDGs}
Do tidal dwarf galaxies contribute significantly to the overall population
of dwarf galaxies ? The answer to this fundamental question relies on
the knowledge of the frequency of tidal interactions between galaxies,
which is still controversial,  and on the life-time
 of TDGs.  The latter
is limited by the hostile environment for TDGs located in the vicinity of 
 giant parent galaxies. They may 
 fall back on their progenitors in time scales of 1~Gyr, as put forward
by Hibbard \& Mihos (1995), or be tidally disrupted. It is therefore expected
that only the most massive TDGs that are far away from their
 progenitors will survive in timescale of Gyrs.
This limits the number of galaxies produced to one or two per colliding
system. In this context, it is not surprising  that no luminous
 star--forming TDGs are found at the 
base of the tidal tails too close from the parent galaxies.

From an observational point of view, the census of TDGs is not an easy 
task. TDGs should obviously be searched in the
environment of  interacting galaxies and a priori in high density 
regions such as groups and clusters. Instances of  small
 star-forming entities were discovered in the arms of the Stephan's
Quintet  by Ohyama et al (1997).
 Hunsberger et al. (1996) claim from the analysis
of photometric data  that half 
of the dwarf galaxies in  Hickson compact groups could be of tidal origin.
However, one should note that  
once the stellar/gaseous bridge between the parent and child
galaxies has dissipated, it is difficult to re-establish a 
 link between the two. Our study has shown that a good genetic
fingerprint of TDGs is their high metallicity.
In this respect, several  studies have put forward trends for
dwarf galaxies in  clusters to be be more metallic than 
field dwarfs (Bothun et al.1985; Vilchez 1995). Since the
collision rate is enhanced in denser environments, it is tempting
to argue that a significant fraction of dwarfs in clusters could
be recycled objects. 
A bimodal star formation history  is also a strong signature for tidal
dwarfs. 
 Evolutionary Synthesis Models simulating a burst of star formation
  on top of the underlying component of old  galaxies
reproduce well the TDG star formation history and will give constraints
for their future evolution (Fritze - v. Alvensleben \& Duc, 1997).\\

\subsection{TDGs as laboratories}
Tidal dwarf galaxies are of particular interest to studies of galaxy formation in the nearby Universe.  They are recently formed galaxies from the collapse
of massive gas clouds. Contrary to BCDGs, TDGs born in pure HI tails are not contaminated
by old stellar populations, allowing a better study of 
the parameters that rule star formation in galaxies.


\begin{thebibliography}{}

%\bibitem[\protect\astroncite{Arp}{1973}]{Arp73}
%{Arp}, H.: 1973,
%\newblock {\em ApJ} {\bf 183}, 411

\bibitem[\protect\astroncite{Barnes}{1992}]{Barnes92}
{Barnes}, J.~E. and {Hernquist}, L.: 1992,
\newblock {\em Nature} {\bf 360}, 715

\bibitem[\protect\astroncite{{Bothun} et~al.}{1985}]{Bothun85}
{Bothun}, G.~D., {Mould}, J.~R., {Wirth}, A., and {Caldwell}, N.: 1985,
\newblock {\em AJ} {\bf 90}, 697

\bibitem[\protect\astroncite{Dubinski}{1996}]{Dubinski96}
Dubinski, J., Mihos, C., and Hernquist, L.: 1996,
\newblock {\em ApJ,} {\bf 462}, 576

\bibitem[\protect\astroncite{Duc}{1994}]{Duc94}
Duc, P.-A and Mirabel, I.~F.: 1994,
\newblock {\em A\&A,} {\bf 289}, 83

\bibitem[\protect\astroncite{Duc}{1995}]{Duc95}
Duc, P.-A.: 1995,
\newblock {\em Ph.D. thesis}, Universit\'e Paris VI

\bibitem[\protect\astroncite{Duc et~al.}{1997}]{Duc97b}
Duc, P.-A., Brinks, E., Wink, J.~E., and Mirabel, I.~F.: 1997,
\newblock {\em A\&A,} {\bf 326}, 537



\bibitem[\protect\astroncite{Duc and Mirabel}{1997}]{Duc97f}
Duc, P.-A. and Mirabel, I.~F.: 1997,
\newblock {submitted to A\&A}

\bibitem[\protect\astroncite{Elmegreen et~al.}{1993}]{Elmegreen93}
Elmegreen, B.~G., Kaufman, M., and Thomasson, M.: 1993,
\newblock {\em ApJ} {\bf 412}, 90

\bibitem[\protect\astroncite{Fritze-v.Alvensleben and Duc}{1997}]{Fritze97}
Fritze-v.Alvensleben, U. and Duc, P.-A.: 1997,
\newblock in {\em IAU JD2-050P}

\bibitem[\protect\astroncite{Hawarden}{1996}]{Hawarden96}
Hawarden, T.G. and Chaytor, D.H.
\newblock {\em BAAS,}{\bf 189}, 120.17


\bibitem[\protect\astroncite{Hibbard et~al.}{1997}]{Hibbard97}
Hibbard, J., van~der Hulst, J., and Barnes, J.: 1997,
\newblock {\em in preparation}

\bibitem[\protect\astroncite{Hibbard et~al.}{1994}]{Hibbard94}
Hibbard, J.~E., Guhathakurta, P., van Gorkom, J.~H., and Schweizer, F.: 1994,
\newblock {\em AJ} {\bf 107}, 67

\bibitem[\protect\astroncite{{Hibbard} and {Mihos}}{1995}]{Hibbard95b}
{Hibbard}, J.~E. and {Mihos}, J.~C.: 1995,
\newblock {\em AJ} {\bf 110}, 140

\bibitem[\protect\astroncite{{Hibbard} and {van Gorkom}}{1996}]{Hibbard96}
{Hibbard}, J.~E. and {van Gorkom}, J.~H.: 1996,
\newblock {\em AJ} {\bf 111}, 655

\bibitem[\protect\astroncite{{Hunsberger} et~al.}{1996}]{Hunsberger96}
{Hunsberger}, S.~D., {Charlton}, J.~C., and {Zaritsky}, D.: 1996,
\newblock {\em ApJ} {\bf 462}, 50

\bibitem[\protect\astroncite{Malphrus et~al.}{1997}]{Malphrus97}
Malphrus, B., Simpson, C., Gottesman, S., and Hawarden, T.~G.: 1997,
\newblock {\em AJ} {\bf 114}, 1427

\bibitem[\protect\astroncite{Mirabel et~al.}{1992}]{Mirabel92}
Mirabel, I.~F., Dottori, H., and Lutz, D.: 1992,
\newblock {\em A\&A} {\bf 256}, L19

\bibitem[\protect\astroncite{Ohyama et~al.}{1997}]{Ohyama97}
Ohyama, Y., Nishiura, S., Murayama, T. and Taniguchi, Y., 1997,
\newblock {\em preprint} 


\bibitem[\protect\astroncite{{Richer} and {McCall}}{1995}]{Richer95}
{Richer}, M.~G. and {McCall}, M.~L.: 1995,
\newblock {\em ApJ} {\bf 445}, 642

\bibitem[\protect\astroncite{Schombert}{1990}]{Schombert90}
{Schombert}, J.~M. and {Wallin}, J.~F. and {Struck-Marcell}, C.: 1990,
\newblock {\em ApJ} {\bf 99}, 497

\bibitem[\protect\astroncite{Schweizer}{1978}]{Schweizer78}
Schweizer, F.: 1978,
\newblock in E. Berkhuijsen and R. Wielebinski (eds.), {\em Structure and
  Properties of Nearby Galaxies}, p. 279, Dordrecht, D. Reidel Publishing Co.

\bibitem[\protect\astroncite{{Smith}}{1994}]{Smith94}
{Smith}, B.~J. and {Higdon}, J.~L.: 1994,
\newblock {\em AJ} {\bf 108}, 837

\bibitem[\protect\astroncite{{Thuan}}{1983}]{Thuan83}
{Thuan}, T.~X.: 1983,
\newblock {\em ApJ} {\bf 268}, 667

\bibitem[\protect\astroncite{{Vilchez}}{1995}]{Vilchez95}
{Vilchez}, J.~M.: 1995,
\newblock {\em AJ} {\bf 110}, 1090

\bibitem[\protect\astroncite{{Wallin}}{1990}]{Wallin90}
{Wallin}, J.~F.: 1990,
\newblock {\em AJ} {\bf 100}, 1477

\bibitem[\protect\astroncite{{Young}}{1991}]{Young91}
{Young}, I. S. and {Scoville}, N. Z.: 1991,
\newblock {\em ARA\&A} {\bf 29}, 581
\end{thebibliography}
\end{document}